\documentclass[%
superscriptaddress,
twocolumn,
groupedaddress,
showpacs,
amsmath,amssymb,
aps,
prl,
floatfix,
showkeys,
reprint,
]{revtex4-1}

\usepackage{graphicx}
\usepackage{dcolumn}
\usepackage{bm}
\usepackage{amsmath}
\usepackage{amssymb}
\usepackage{multirow}
\usepackage{booktabs}
\usepackage{tabularx}
\usepackage[english]{babel}
\usepackage[utf8]{inputenc}
\usepackage{enumerate}
\usepackage{newunicodechar}
\newunicodechar{ﬁ}{fi}
\newunicodechar{ﬀ}{ff}
\usepackage{gensymb}

\newcommand{\etal}{\textit{et al.\/}}
\newcommand{\ie}{i.\,e.,}

\begin{document}

\title{Noncollinear spin density of an adatom on a magnetic surface}

\author{Soumyajyoti Haldar}
\email[Corresponding author: ]{haldar@physik.uni-kiel.de}
\affiliation{Institute of Theoretical Physics and Astrophysics, University of Kiel, Leibnizstrasse 15, 24098 Kiel, Germany}

\author{Stefan Heinze}
\affiliation{Institute of Theoretical Physics and Astrophysics, University of Kiel, Leibnizstrasse 15, 24098 Kiel, Germany}

\date{\today}

\begin{abstract}
	We show that an individual adatom on a magnetic surface can exhibit a noncollinear spin density.
	Using density functional theory we study Co and Ir adatoms on a Mn monolayer on the W(110) surface which possesses 
	a noncollinear canted spin structure.
	Due to hybridization with the nearest and next-nearest Mn atoms of the monolayer the spin direction of the underlying substrate
	is encoded into the orbitals of the adatom. 
	This explains recent scanning tunneling microscopy experiments showing a spin sensitive shape asymmetry of adatoms [Serrate {\etal}, Phys.~Rev.~B {\bf 93}, 125424 (2016)]
	which confirms the intra-atomic noncollinear magnetism of the adatoms.
\end{abstract}

\maketitle

Noncollinear magnetic structures at surfaces and interfaces are being intensively studied today
due to their intriguing dynamical and transport properties making them promising candidates for spintronic
applications \cite{Fert2013,Nagaosa2013}. Prominent examples are spin spirals \cite{Bode2007,Ferriani2008,Phark2014}, chiral domain walls \cite{Heide2008,Ryu2013,Emori2013}
and skyrmions~\cite{Bogdanov1989,Bogdanov:2001aa,Heinze2011,Romming2013,Herve2018}. In these spin structures the direction of the magnetic moment changes 
from one atomic site to the next. 
A signature of this inter-atomic noncollinearity is also found in the electronic structure~\cite{Sandratskii1998} which can be
locally resolved by scanning tunneling microscopy as recently shown \cite{Hanneken2015,Crum2015,Fischer2016}.

As first predicted by
Nordstr\"om {\etal}~\cite{Nordstrom1996} there is in addition intra-atomic noncollinear magnetism in which the magnetization
direction varies within the orbitals of a single atom. It can be induced by spin-orbit coupling \cite{Nordstrom1996} or by a
noncollinear magnetic structure~\cite{Oda1998}. This effect has been intensively studied based on first-principles 
electronic structure theory~\cite{Nakamura2003,Nakamura2004a,Nakamura2004b,Sharma2007}. 
However, direct experimental detection is difficult. Spin-polarized scanning tunneling microscopy (SP-STM) experiments provided
evidence for intra-atomic noncollinear magnetism present in the apex atom of the tip~\cite{Bode2001} and of atoms in a reconstructed
surface alloy \cite{Gao2008}. Recently, SP-STM experiments by 
Serrate {\etal}~\cite{Serrate2010,Serrate2016} showed that the image of an individual Co adatom on a Mn monolayer on W(110)
becomes asymmetric if the tip magnetization is at an angle with that of the Co adatom. A noncollinear
spin density above the Co adatom can be inferred from this experiment~\cite{Serrate2016}. However, the phenomenological model introduced in Ref.~\cite{Serrate2016}
requires a continuous spin rotation by an angle which is an order of magnitude larger than that of the underlying spin spiral
and does not provide a relation to the electronic states of the Co atom.

Here, we demonstrate based on density functional theory (DFT) that the spin density of an adatom on a magnetic surface with a
noncollinear spin structure obtains the canted spin directions within its orbitals. We consider Co and Ir adatoms on Mn/W(110) as
an example and show that by hybridization the spin structure of the nearest and next-nearest Mn atoms is encoded into different $d$ orbitals of the adatoms. 
This allows a direct imaging of the noncollinear spin density of the adatom by SP-STM. Our calculations explain the experiments
of Serrate {\etal}~\cite{Serrate2016} which confirms the presence of intra-atomic noncollinear magnetism in such adatoms.

\begin{figure}
	\centering
	\includegraphics[scale=0.9,clip]{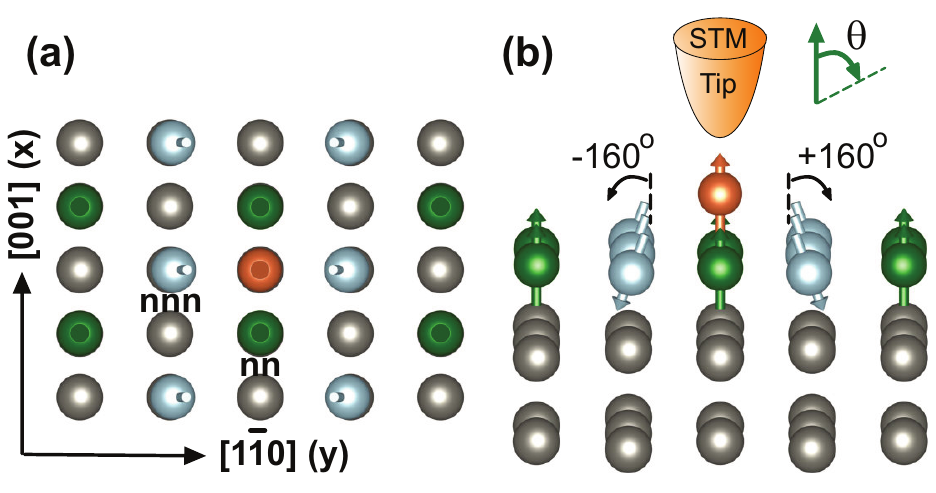}
	\caption{(a) Top view and (b) perspective view of the supercell along with schematics of an SP-STM tip with a magnetization direction
  at an angle of $\theta$ with respect to the magnetic moment of the adatom. Gray spheres represent W atoms while Mn atoms are depicted as green (light blue) spheres with arrows
  showing the noncollinear magnetic moments ($\pm 160\degree$) with respect to the magnetic moment of the adatom (orange sphere). The nearest neighbor and next nearest neighbor Mn atoms 
  to the adatom are labeled nn and nnn, respectively. $x$ and $y$ refer to the notation used for labeling of the $d$ orbitals. }
	\label{fig:geom}
\end{figure}

We used DFT within the projector augmented-wave method (PAW)~\cite{blo,blo1} as implemented in the \textsc{vasp} code~\cite{vasp1,vasp2} to study the electronic and magnetic properties of Co and Ir adatoms on Mn/W(110). The generalized gradient
approximation (GGA) of Perdew-Burke-Ernzerhof
(PBE) is used for the exchange-correlation~\cite{PBE,PBEerr}. A 450 eV energy cutoff is used for the plane wave basis set. 
Noncollinear magnetism is taken into account as described in~\cite{Hobbs2000}. We used 144 k$_\parallel$-points in the full two-dimensional Brillouin zone for the calculation of noncollinear electronic properties. 
\begin{figure*}[htbp]
	\centering
	\includegraphics[scale=0.9,clip]{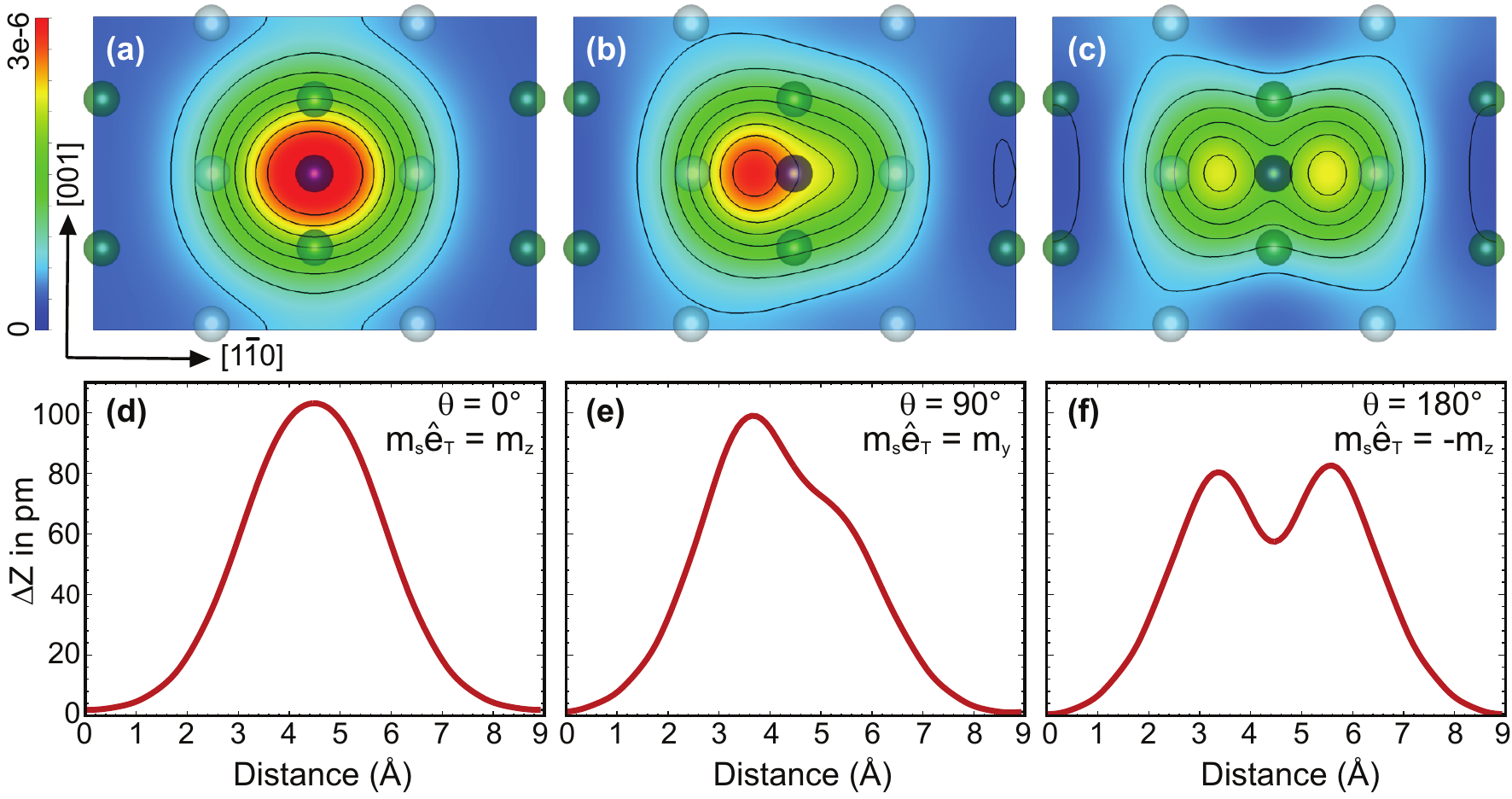}
	\caption{Simulated SP-STM images [(a)$-$(c)] in the vacuum 3 {\AA} above a Co adatom adsorbed onto a Mn/W(110) substrate within the energy range ($E_F$, $E_F+0.1$ eV) for three different tip-atom magnetization orientation ($\theta$ = 0$\degree$, 90$\degree$, and 180$\degree$) showing gradual development of the shape asymmetry. The positions of Co and Mn atoms are marked for clarity. Corresponding line profiles [(d)$-$(f)] across the center of the Co adatom along $[1\overline{1}0]$ in [(a)$-$(c)]. Corrugation amplitudes have been obtained according to Ref.~\cite{Heinze2006}.}
	\label{fig:stm}
\end{figure*} 
The system is modeled using a symmetric slab consisting
of five atomic layers of W with a pseudomorphic Mn layer on each side, as has been found experimentally~\cite{Bode1998}. A thick vacuum layer of $\approx$ 25 {\AA} is included in the direction normal to the surface to ensure no spurious interactions between repeating slabs. We used a $c (4\times4)$ 
surface unit cell, as shown in Fig.~\ref{fig:geom}(a), with the GGA lattice constant of W, {\ie} 3.17 {\AA}. The adatom was added on each Mn layer at the hollow-site position with a minimum 6.34 {\AA} distance apart from its periodic image to keep the interactions between them negligible~\cite{Caffrey2014a}. 

We apply the constrained local moments approach where the magnetic moments of nearest-neighbor Mn atoms are kept parallel to the magnetic moment of the adatom 
and the magnetic moments of the next nearest-neighbor Mn atoms are pointing opposite of the adatom with a relative angle of $\pm 160\degree$ between them (see Fig.~\ref{fig:geom}). 
This magnetic structure is locally as the full spin spiral ground state \cite{Bode2007} in which the magnetic moments of adjacent Mn rows rotate by a
constant angle~\footnote{Note that we use a canting angle slightly larger than that in the experimentally observed spin spiral of
about $170\degree$ which leads to qualitatively the same SP-STM image [Fig.~\ref{fig:stm}(b)] for a canted tip magnetization but 
a larger asymmetry of the line profile [Fig.~\ref{fig:stm}(e)] than observed experimentally~\cite{Serrate2016}.}.
Therefore, the spin-direction dependent hybridization of the Co states with that of the Mn neighbors is captured within our calculation.

First we discuss how the noncollinear magnetism of the Co adatom is probed by the magnetic tip of a scanning tunneling microscope (see Fig.~\ref{fig:geom}(b)).
To simulate SP-STM images from our DFT calculations which can be directly compared with the experiments of Serrate {\it et al.} \cite{Serrate2016} 
we used the extension of the Tersoff-Hamann model~\cite{Tersoff1983,Tersoff1985} to SP-STM~\cite{Wortmann2001,Heinze2006}.
For the sake of simplicity we consider a small bias voltage. The tunneling current as a function of tip position
$\bm{R}_{\rm T}$ is then given by:
\begin{equation}
I(\bm{R}_{\rm T}) \propto n_{\mathrm{S}}(\bm{R}_{\rm T},E_{\rm F}) + P_{\mathrm{T}} \bm{\hat{e}}_\mathrm{T} \bm{m}_\mathrm{S}(\bm{R}_{\rm T},E_{\rm F})
\label{eq:current}
\end{equation}
where $n_{\mathrm{S}}(\bm{R}_{\rm T},E_{\rm F})$ and $\bm{m}_\mathrm{S}(\bm{R}_{\rm T},E_{\rm F})$ are the local density of states (LDOS) and the
local magnetization density of states of the sample at the Fermi energy $E_{\rm F}$, respectively. The spin polarization of the tip, 
$P_\mathrm{T}$ 
is typically taken as 0.5 and $\bm{\hat{e}}_\mathrm{T}$ is the unit vector along the tip magnetization. As a result, the SP-STM tip is 
sensitive only to the projection of $\bm{m}_\mathrm{S}$ onto its magnetization direction.

For a noncollinear magnetic sample the
electronic states can be described by two-component spinors 
\begin{equation}
\Psi_\mu (\bm{r}) =\begin{pmatrix}
\psi_{\mu\uparrow} (\bm{r}) \\\psi_{\mu\downarrow} (\bm{r}) \end{pmatrix}\,.
\end{equation}
The local magnetization density of states at energy $\epsilon$ is given by 
\begin{eqnarray}
{\mathbf{m}}({\mathbf R}_{\scriptscriptstyle T},\epsilon) &=&
\sum\limits_{\mu} \delta (\epsilon_\mu-\epsilon)
\Psi_\mu^{\dagger}({\mathbf R}_{\scriptscriptstyle T}) \,
\boldsymbol{\sigma} \, \Psi_\mu({\mathbf R}_{\scriptscriptstyle T})
\label{eq:SLDOS}
\end{eqnarray} 
where $\epsilon_\mu$ is the energy of $\Psi_\mu$. The LDOS of the sample ${n}_S ({\mathbf R}_{\scriptscriptstyle T},\epsilon)$ is obtained by
Eq.~(\ref{eq:SLDOS}) replacing Pauli's spin matrix $\boldsymbol{\sigma}$ by the unit matrix.

For a single state $\Psi_\mu$ the three magnetization components can be written explicitly as
\begin{eqnarray}
m^x_{\mu}(\bm{r}) & = & 
\psi^*_{\mu\uparrow}(\bm{r})\psi_{\mu\downarrow}(\bm{r}) + 
\psi^*_{\mu\downarrow}(\bm{r})\psi_{\mu\uparrow}(\bm{r})
\label{eq:mx}\\
m^y_{\mu}(\bm{r}) & = & 
-i [\psi^*_{\mu\uparrow}(\bm{r})\psi_{\mu\downarrow}(\bm{r}) - 
\psi^*_{\mu\downarrow}(\bm{r})\psi_{\mu\uparrow}(\bm{r})]
\label{eq:my}\\
m^z_{\mu}(\bm{r}) & = & 
|\psi_{\mu\uparrow}(\bm{r})|^2 - |\psi_{\mu\downarrow}(\bm{r})|^2.
\label{eq:mz}
\end{eqnarray} 
For a spinor function with non-zero spin-up and spin-down components, there is also a non-zero magnetization component orthogonal to the quantization axis. 
This is a key difference of a noncollinear vs.~a collinear magnetic structure. 

\begin{figure*}
	\centering
	\includegraphics[scale=0.9,clip]{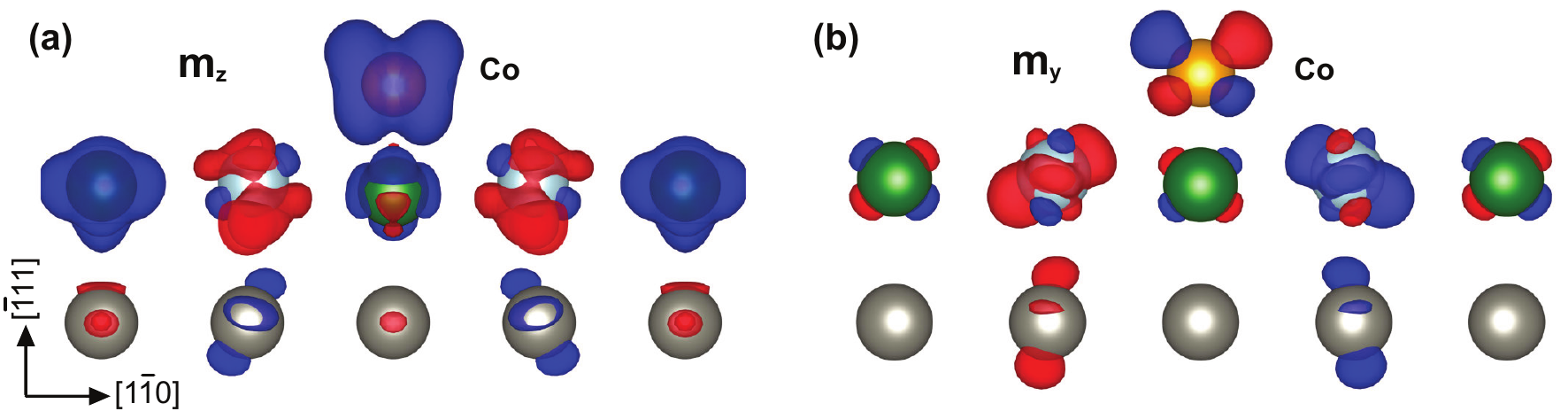}
	\caption{Plot of the calculated local magnetization density of states of a Co adatom on Mn/W(110) 
  integrated over the energy range ($E_F$, $E_F+0.1$ eV). 
  (a) and (b) panels show $m_z$ and $m_y$ components, respectively.
  Red (blue) denotes positive (negative) values of the isosurfaces. The isosurfaces are plotted with $1/10^{th}$ of the maximum isosurface values. Gray spheres indicate W atoms, green (light blue) spheres
  Mn atoms and orange sphere denotes the Co atom.}
	\label{fig:parchg-co}
\end{figure*}

Figures~\ref{fig:stm}(a-c) 
show simulated constant-height SP-STM images 
of a Co adatom on Mn/W(110) 
at a bias voltage of $+0.1$~V and for three different angles between the tip and adatom magnetization directions. 
For a tip magnetization parallel to the Co magnetic moment ($\theta$ = $0\degree$) 
the second, spin polarized term in the tunneling current, Eq.~(\ref{eq:current}), is the $z$-component of the local magnetization density. 
The STM image shows a nearly circularly symmetric structure centered at the position of the Co adatom [Fig.~\ref{fig:stm}(a)] 
and the corresponding line profile is mirror symmetric with respect to the Co position [Fig.~\ref{fig:stm}(d)]. 
As shown previously assuming an antiferromagnetic collinear state of the Mn layer~\cite{Serrate2010} the image originates 
from the spin-up component which is of $s$ and $p_z$ orbital character.
For an opposite tip magnetization direction ($\theta=180\degree$) the image shows a two lobe structure [Fig.~\ref{fig:stm}(c)]
due to the spin-down component which is of $d_{yz}$ type in agreement with Ref.~\cite{Serrate2010}. 
The corresponding line profile [Fig.~\ref{fig:stm}(f)] shows two peaks on the two sides of the adatom's position and a valley exactly at the center 
position of the adatom~\footnote{Ideally, the two lobes in the STM image [Fig.~\ref{fig:stm}(c)] and the corresponding line profile [Fig.~\ref{fig:stm}(f)] should be perfectly symmetric. However, a very small asymmetry is observed in our DFT results due to numerical inaccuracies.}.

If the tip magnetization is perpendicular to the Co magnetic moment ($\theta$ = $90\degree$) 
the $y$-component of the local magnetization density of states
appears as the second term of Eq.~(\ref{eq:current}). 
If all states $\Psi_\mu$ were collinear, i.e.~had only a spin-up or a spin-down component, $m^y_{\mu}(\bm{r})$ would vanish according to Eq.(\ref{eq:my})
and the spin-polarized term in the tunneling current, Eq.~(\ref{eq:current}), would be zero, i.e.~the STM image would be given by the local density of states.
In that case 
an arbitrary angle $\theta$ between tip and adatom magnetization
would result in a superposition of the images Fig.~\ref{fig:stm}(a) and (c) and always possess a mirror symmetric line profile.
In contrast the STM image which we obtain from our DFT calculations including the noncollinearity of the Mn spin structure [Fig.~\ref{fig:stm}(b)]
is asymmetric with the left lobe becoming much stronger than the right lobe. As a result there is a maximum in the STM image clearly shifted away from
the position of the Co adatom.     
The line profile [Fig.~\ref{fig:stm}(e)] now lacks the mirror symmetry.
Both the obtained SP-STM image and the line profile are in excellent agreement with the observations of Serrate {\it et al.} \cite{Serrate2010,Serrate2016}. 

In order to understand the origin of these asymmetries in the SP-STM images and in the line profiles, we have analyzed the 
magnetization densities for the Co adatom on Mn/W(110). Since the magnetic moments of all atoms are in the $yz$-plane the $m_x$ component vanishes by symmetry.
In the $m_z$ component [Fig.~\ref{fig:parchg-co}(a)] $d_{\downarrow yz}$ and $d_{\downarrow xy}$ orbitals of the Co adatom contribute within the considered
energy window ($E_F$, $E_F+0.1$ eV)(see Fig.~S1 in Supplemental Material for projected density of states.).  
This is reflected in Fig.~\ref{fig:parchg-co}(a), where a pronounced negative isosurface can be observed in the vicinity of the Co adatom as expected from the 
spin-down component, cf.~Eq.~(\ref{eq:mz}). The two lobe structure of this isosurface above the atom 
is responsible for the SP-STM image for $\theta=180\degree$ [Fig.~\ref{fig:stm}(c)]
and leads to the mirror symmetric line profile [Fig.~\ref{fig:stm}(f)].
The positive part of the isosurface which leads to the circularly symmetric image [Fig.~\ref{fig:stm}(a)] is not visible in Fig.~\ref{fig:parchg-co}(a) since it stems
from $s,p_z$ states that are more delocalized and only contribute significantly farther from the Co adatom (see Fig.~S2 in the Supplemental Material).

For the  $m_y$ component [Fig.~\ref{fig:parchg-co}(b)] we observe a four lobe structure at the Co adatom, however, with both positive and negative sign of the isosurface. 
A strong hybridization with the next-nearest neighbor Mn atoms is visible. 
Thereby, the positive $m_y$ value of the left Mn atom is transferred 
to the right lobe above the Co adatom and vice versa for the negative component.
Thus the sign of the $m_y$ component in the vacuum which is detected by the STM tip is opposite to that from the underlying Mn atoms. In terms of the spinor wave functions the
state at the Co atom exhibits a $d_{yz}$ character in the $\psi_\downarrow$ component while
it is a superposition of $s$, $d_{x^2-y^2}$, and $d_{xy}$ type in the $\psi_\uparrow$ component. According to Eq.~(\ref{eq:my})
this leads to an $m_y$ component which has altering positive and negative signs in the four lobes as the $d_{yz}$ orbital. 

By combining the information on the $m_y$ and the $m_z$ component with the corresponding orbitals
we conclude that the spin direction in the vacuum density above the Co atom 
changes from pointing to the lower left, to pointing upwards, and finally pointing to the lower right as we move from left to right across the atom. 
This is in striking contrast to the continuous spin rotation above the adatom previously assumed~\cite{Serrate2016}.

\begin{figure*}
	\centering
	\includegraphics[scale=0.9,clip]{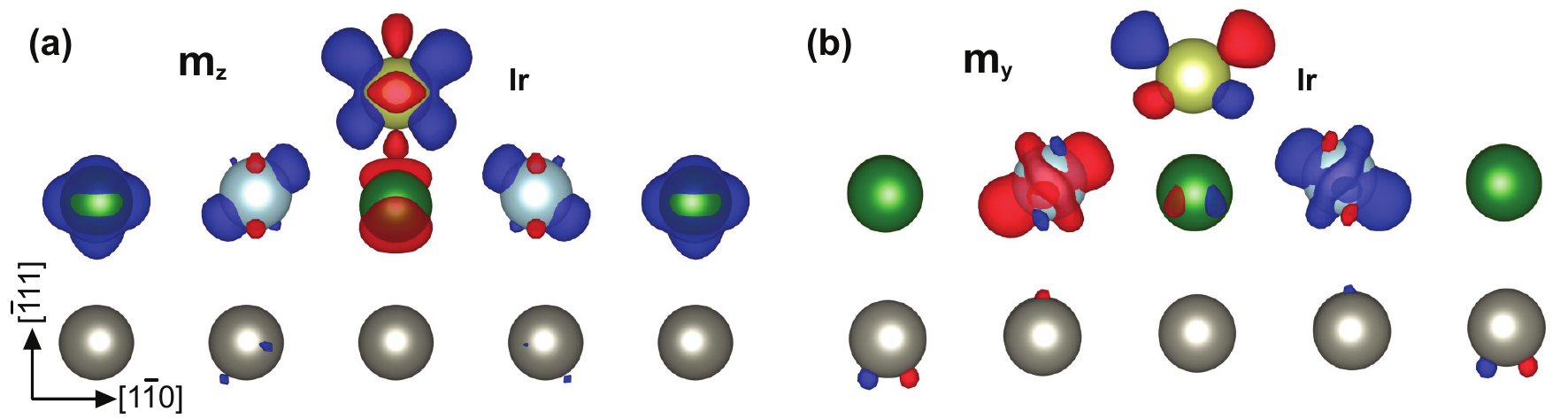}
	\caption{Plot of the calculated local magnetization density of states of an Ir adatom on Mn/W(110) 
		integrated over the energy range ($E_F-0.1$~eV, $E_F$). 
		(a) and (b) panels show $m_z$ and $m_y$ components, respectively.
		Red (blue) denotes positive (negative) values of the isosurfaces. The isosurfaces are plotted with $1/10^{th}$ of the maximum isosurface values. Gray spheres indicate W atoms, green (light blue) spheres
		Mn atoms and orange sphere denotes the Ir atom.}
	\label{fig:parchg-ir}
\end{figure*}

We have also investigated the effect of spin orbit coupling (SOC) on the noncollinear magnetic structure of the Co adatom on Mn/W(110). SOC is taken into account 
as described in Ref.~\cite{Hobbs2000}. Our calculations show that the effect of SOC in this system is weak \cite{Caffrey2014a} 
as expected for a $3d$ transition-metal such as Co. In particular, it does not lead to the spin-mixing responsible
for the orbital dependent spin direction discussed above.

To show that intra-atomic noncollinear magnetism can be observed also for other adatoms we have calculated the magnetization density
for an Ir adatom on Mn/W(110).
At the Fermi energy the Ir atom exhibits a peak from $d_{yz}$ states in the spin-down channel as Co \cite{Caffrey2014a}.  
In the $m_z$ component [Fig.~\ref{fig:parchg-ir}(a)] contributions from $d_{\downarrow yz}$, $d_{\downarrow xy}$, and $d_{\uparrow z^2}$ orbitals of the Ir adatom can be observed.
Above the Ir atom this leads to a positive isosurface lobe from the $d_{\uparrow z^2}$ orbital sticking out between the two negative isosurface lobes from the
$d_{\downarrow yz}$ states as seen in Fig.~\ref{fig:parchg-ir}(a). 
The $d_{\downarrow yz}$ orbitals of the Ir atom hybridize with rotated $d_{\downarrow z^2}$ states of the next-nearest neighbor Mn atoms which is visible in both
magnetization components [Fig.~\ref{fig:parchg-ir}(a,b)]. The $m_y$ component of the Ir adatom [Fig.~\ref{fig:parchg-ir}(b)] is as that of Co and caused by similar 
spinor wave functions. Due to the $d_{\uparrow z^2}$ state at the Fermi energy the encoding of the spin direction into the Ir $d$ orbitals is here more 
obvious than for the Co adatom.
 
In conclusion we have solved the puzzle of the spin sensitive 
shape asymmetry observed in SP-STM experiments of Co adatoms on Mn/W(110)~\cite{Serrate2010,Serrate2016}. We demonstrated that the origin is the noncollinear 
spin density of the Co adatom. It is induced by the hybridization with the orbitals of the Mn atoms which encodes their canted spin structure
into different orbitals of the adatom. This effect applies generally for adatoms on surfaces with a noncollinear magnetic structure. 
Our study shows that the experiment of Serrate {\it et al.}~\cite{Serrate2010,Serrate2016}
can be seen as a direct confirmation of intra-atomic noncollinear magnetism of an adatom. 
 
We acknowledge the DFG via SFB677 for financial support. We gratefully acknowledge the computing time at the supercomputer of the North-German Supercomputing 
Alliance (HLRN).

%

\end{document}